\documentclass[conference]{IEEEtran}
\IEEEoverridecommandlockouts
\usepackage{cite}
\usepackage{amsmath,amssymb,amsfonts}
\usepackage{algorithmic}
\usepackage{graphicx}
\usepackage{textcomp}
\usepackage{xcolor}

\usepackage{booktabs}
\usepackage{array}
\usepackage{url}
\usepackage{listings}
\usepackage{xcolor}
\usepackage{subcaption}
\usepackage{tcolorbox}
\usepackage{enumitem}
\usepackage{tabularx}
\usepackage[para]{footmisc}
\def\BibTeX{{\rm B\kern-.05em{\sc i\kern-.025em b}\kern-.08em
    T\kern-.1667em\lower.7ex\hbox{E}\kern-.125emX}}
\begin{document}

\makeatletter
\def\ps@IEEEtitlepagestyle{%
  \def\@oddfoot{}%
  \def\@evenfoot{}%
  \def\@oddhead{%
    \parbox[b]{\textwidth}{\centering
    Author copy of paper published at \textit{17th International Conference on Cloud Computing Technology and Science \\(IEEE CloudCom2026)}}
  }%
  \def\@evenhead{\@oddhead}%
}
\makeatother

\title{AgentWare: Automating the Lifecycle of Agentic Applications across the Edge-to-Cloud Continuum\\
}

\author{
\IEEEauthorblockN{Michalis Kasioulis, Moysis Symeonides,  George Pallis, Marios D. Dikaiakos}
\IEEEauthorblockA{
Department of Computer Science, University of Cyprus\\
\{mkasio01, msymeo03, pallis, mdd\}@ucy.ac.cy
}
}


\maketitle

\begin{abstract}
Deploying LLM-enabled agentic applications across the Edge-to-Cloud continuum remains challenging due to hardware heterogeneity, deployment complexity, limited observability, and the lack of systematic evaluation methods. Existing solutions address agent development, observability, or benchmarking separately, offering limited support for the full lifecycle of distributed agentic applications.
This paper presents \textit{AgentWare}, an AgenticOps framework that automates the provisioning, deployment, observability, and evaluation of agentic applications across Edge-to-Cloud infrastructures. AgentWare introduces an end-to-end lifecycle pipeline that automatically prepares heterogeneous execution environments, transforms user-defined agent implementations into distributed applications, deploys agent components across the continuum, and performs unified collection of execution traces, infrastructure telemetry, and evaluation metrics. The framework further supports automated semantic evaluation through LLM-as-a-Judge workflows and generates reproducible reports covering correctness, performance, resource utilization, and energy consumption.
We demonstrate the applicability of AgentWare through a distributed book assistant agent deployed across real Edge-to-Cloud infrastructure under multiple deployment and model configurations. The results show that AgentWare enables systematic experimentation and evaluation of distributed agentic applications while significantly reducing the manual effort required for deployment, instrumentation, and analysis.
\end{abstract}


    \vspace*{-.2\baselineskip}
\section{Introduction}
    \vspace*{-.2\baselineskip}

\noindent Large Language Models (LLMs) have enabled a new generation of intelligent applications capable of autonomously performing complex tasks across diverse domains. Commonly referred to as \emph{agentic AI}, these systems augment LLMs with memory, external tools, and execution capabilities, allowing them to reason, plan, and interact with their environment to achieve user-defined goals with limited human intervention.
Agentic AI systems are rapidly being adopted in applications ranging from personal assistants and customer support to scientific discovery and industrial automation~\cite{fourney-2024}.
Moreover, the increasing availability of sensing and control interfaces further motivates the deployment of agentic applications across the Edge-to-Cloud continuum, where LLMs and tools are distributed over heterogeneous devices with varying computational capabilities and resource constraints~\cite{al-2024}.

However, deploying agentic applications in such environments introduces several challenges. Hardware heterogeneity complicates portability and reproducibility~\cite{belcastro-2025}, while distributed execution increases orchestration complexity and sensitivity to resource availability.
Moreover, observability requires correlating agent-level execution traces with infrastructure-level metrics~\cite{moshkovich-2025}. 
Existing approaches remain limited, focusing primarily on accuracy and latency while overlooking semantic correctness, safety, and energy consumption, all of which are critical for real-world cyber-physical deployments~\cite{mohammadi-2025}.
To address these challenges, existing systems span three largely disconnected directions: agent development frameworks 
(e.g., Smolagents\footnote{\url{huggingface.co/smolagents}}, LangChain\footnote{\url{langchain.com}}), observability stacks (e.g., Phoenix\footnote{\url{arize.com/phoenix}}), and benchmarking suites (e.g., AgentBench~\cite{liu-2023}). 
Although each addresses part of the agent lifecycle, none provides a unified solution for building, deploying, instrumenting, and evaluating Edge-to-Cloud agentic apps.

In this paper, we present \textbf{AgentWare}, a framework enabling AgenticOps across the Edge-to-Cloud continuum. Our contribution is threefold: (i)~we introduce the \textbf{AgentWareOps (AWOps) lifecycle pattern}, which captures the end-to-end process of building, deploying, instrumenting, and evaluating agentic applications in heterogeneous environments; (ii)~we implement this pattern in an \textbf{open-source framework} named AgentWare~\cite{agentware}, which automates the configuration-driven build--deploy--instrument--evaluate pipeline while supporting deployment configuration, LLM selection, and unified collection of performance, accuracy, and energy metrics; and (iii)~we demonstrate its applicability through a \textbf{distributed book assistant agent} deployed across real Edge-to-Cloud infrastructure under multiple deployment and LLM configurations. To the best of our knowledge, AgentWare is among the first frameworks to systematically integrate lifecycle automation, observability, and evaluation support for agentic applications across heterogeneous Edge-to-Cloud environments.

    \vspace*{-.2\baselineskip}
\section{Edge-enabled Agents \& Challenges}
    \vspace*{-.2\baselineskip}

\begin{figure}[t]
\centering
\includegraphics[scale=0.55, trim={0 9 0 0.2cm}, clip]{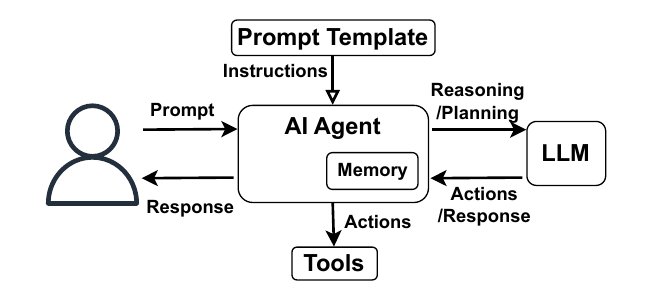}
\caption{AI Agent Architecture} \label{ai-agent-architecture}
    \vspace*{-1.5\baselineskip}
\end{figure}

Unlike traditional applications that follow predefined execution paths, LLM-enabled agents dynamically reason about user requests and environmental context to determine appropriate actions~\cite{loven-2025}. As illustrated in Fig.~\ref{ai-agent-architecture}, an agent is typically composed of four core components: (i) a prompt template defining its role and operational constraints, (ii) an LLM acting as the reasoning engine, (iii) a memory component for retaining contextual information, and (iv) a set of tools that provide access to external services, sensors, databases, APIs, or inference functions.
At runtime, the agent combines the user request, prompt template, memory, and available tool descriptions into a context passed to the LLM. The LLM then determines whether tools should be invoked, updates its reasoning based on returned results, and iteratively executes actions until a final response is generated.
Within the Edge-to-Cloud continuum, agent components can be deployed as distributed services, similarly to cloud-native microservices. Tools may expose sensing, inference, retrieval, or actuation capabilities, enabling closed perception--decision--action loops spanning heterogeneous infrastructures. 
This distributed deployment model supports flexible placement of agent components according to latency, privacy, and resource constraints while maintaining end-to-end agent behavior.

Thus, deploying agentic applications across the Edge-to-Cloud continuum introduces challenges inherited from distributed systems and edge computing, further amplified by the dynamic and non-deterministic nature of LLM-enabled agents. We summarize the most important challenges below:

\noindent \textbf{Hardware Heterogeneity:}
The continuum comprises highly heterogeneous devices ranging from resource-constrained edge nodes to cloud datacenters. These devices differ in computational capabilities, energy profiles, hardware cost, and processor architectures, requiring support for architecture-specific binaries and dependency management across heterogeneous environments~\cite{chen-2020,belcastro-2025}. Furthermore, AI accelerators (e.g., GPUs, TPUs, NPUs) provide workload-specific performance benefits but often expose limited or opaque resource utilization metrics, complicating performance analysis and optimization. As a result, deploying agentic applications consistently across heterogeneous infrastructures remains a significant challenge.

\noindent \textbf{Parameterization and tool placement of Agentic Apps:}
Unlike traditional microservice applications, agentic systems introduce additional configuration dimensions, including LLM selection, prompt design, tool configuration, and inference parameters~\cite{mohammadi-2025}. These choices directly affect task accuracy, latency, resource utilization, and energy consumption. Moreover, Agentic applications consist of multiple components, including LLMs, memory services, and tools that may be distributed across the continuum. Determining their optimal placement is challenging because it depends on the objective being optimized, such as latency, accuracy, privacy, cost, or sustainability. Current agentic frameworks largely assume single-host execution and provide limited support for distributed deployment~\cite{belcastro-2025}. 
As a result, developers must combine agentic frameworks with distributed execution technologies while manually managing orchestration across heterogeneous devices, making systematic experimentation and reproducible deployment workflows essential for evaluating alternative configurations and optimizing agent behavior.

\noindent \textbf{Instrumentation \& Observability:}
Understanding the behavior of distributed agentic applications requires observability across multiple layers, including the application, virtualization, and hardware layers. Agent execution directly influences infrastructure resource utilization, response latency, and energy consumption, requiring the correlation of agent-level traces with system-level telemetry. In addition, agentic observability requires capturing LLM calls, tool invocations, reasoning steps, and execution traces, where each user prompt corresponds to a trace composed of multiple spans.
Collecting, synchronizing, and analyzing information across heterogeneous devices remains a non-trivial task because agent traces and infrastructure metrics are produced by different services and abstraction layers, requiring synchronizing and correlating these heterogeneous data sources to attribute resource consumption to specific agent actions for meaningful analysis.

\noindent \textbf{Evaluation:}
Evaluating edge-enabled agentic applications extends beyond traditional accuracy and latency metrics. Human evaluation does not scale with increasing workload volumes, while semantic evaluation techniques such as LLM-as-a-Judge require collecting and preprocessing execution traces, tool invocations, and intermediate reasoning steps~\cite{zheng-2023}. 
Consequently, evaluation becomes multi-dimensional, requiring the joint assessment of correctness, task success, policy compliance, performance, 
and energy consumption across repeated executions in the presence of non-deterministic agent behavior.


\vspace{-.2\baselineskip}
\section{The AgentWare Framework}
\vspace{-.2\baselineskip}


\begin{figure*}[t]
\centering
\includegraphics[width=0.95\linewidth]{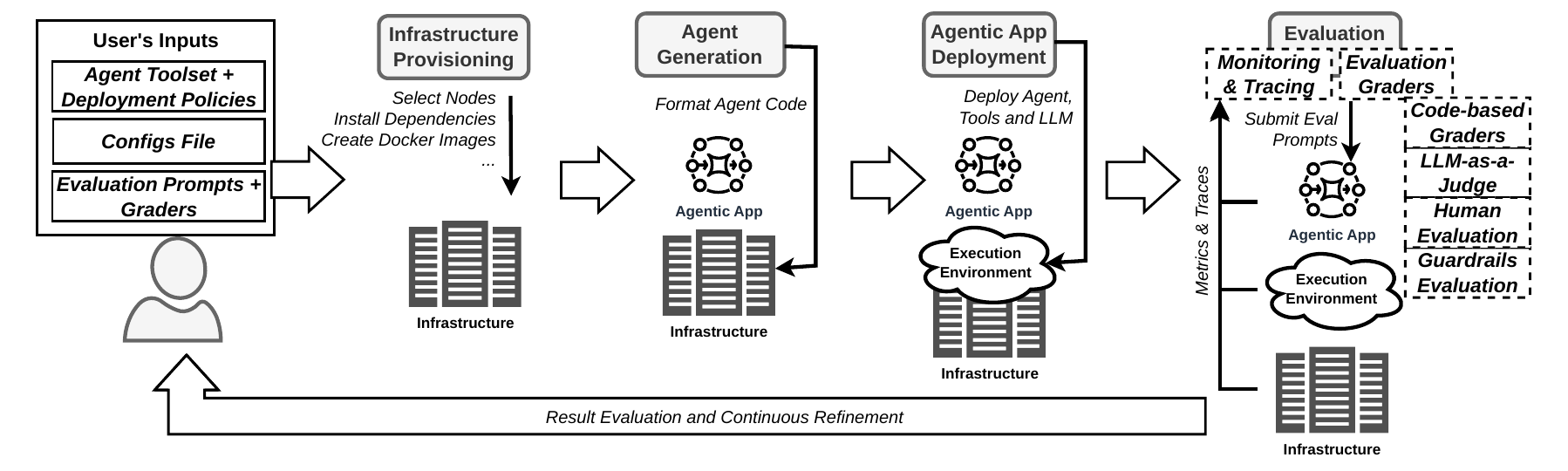}
    \vspace*{-1.\baselineskip}
\caption{AWOps Full-lifecycle Pipeline} \label{fig:pipeline}
    \vspace*{-1.2\baselineskip}
\end{figure*}

Motivated by the challenges discussed above, we propose \textit{AgentWareOps (AWOps)}, a systematic methodology for automating the lifecycle of agentic applications across the Edge-to-Cloud continuum. Inspired by DevOps practices, the AWOps pipeline consists of four stages (Fig.~\ref{fig:pipeline}): \textbf{Infrastructure Provisioning}, \textbf{Agent Generation}, \textbf{Agentic Application Deployment}, and \textbf{Evaluation}.
The pipeline is driven by three user-provided inputs: (i) an \textit{Agentic Toolset} containing the implementation of the agent's tools and associated deployment policies, (ii) a \textit{configuration file} specifying the parameters required by the AWOps pipeline,
and (iii) optionally, a \textit{set of prompts} to be submitted to the agent during the evaluation.

The \textbf{Infrastructure Provisioning} stage addresses \textit{hardware heterogeneity} by preparing the target infrastructure for execution across heterogeneous Edge-to-Cloud environments. This stage abstracts infrastructure-specific differences and establishes a consistent execution foundation despite variations in hardware architectures, accelerators, and resource capabilities.

The \textbf{Agent Generation} stage addresses the challenge of \textit{parameterization}. During this stage, the framework transforms the developer-defined tools implementation code into a deployment-ready distributed application through automated code generation. Based on the specified deployment configuration provided by the user with the tools' implementation code, 
the generated code incorporates the necessary runtime abstractions required for distributed execution across the continuum. By automatically translating high-level deployment specifications into executable artifacts, this stage eliminates manual code modifications and enables reproducible experimentation with alternative placement and parameterization strategies.

The \textbf{Agentic Application Deployment} stage addresses the challenges of \textit{placement} and \textit{observability}. The framework deploys the generated agentic application and instantiates the distributed execution environment across the continuum by automating the deployment of the agent's services in accordance with the specified deployment configuration. As a result, developers can evaluate alternative deployment strategies without manually configuring individual nodes, while maintaining scalable and reproducible execution across heterogeneous infrastructures. In addition, this stage establishes \textit{observability} by automatically deploying and configuring the monitoring and tracing services required to collect agent-level execution traces and infrastructure-level telemetry.

Finally, the \textbf{Evaluation} stage addresses the challenge of \textit{evaluation}. Using the user-provided evaluation prompts and configuration, the framework leverages the telemetry and execution traces collected during deployment to perform multi-dimensional assessment of the agentic application. This includes deterministic grading of objective properties through configurable code-based graders, automated semantic evaluation through a configuration-driven LLM-as-a-Judge grader, and guardrail compliance verification against predefined safety and operational constraints. By combining application-level execution traces with infrastructure-level telemetry, the framework enables systematic and reproducible comparison of deployment configurations while supporting comprehensive analysis of correctness, performance, resource utilization, energy consumption, and policy compliance. 

\vspace*{-.2\baselineskip}
\section{Implementation Details}
\vspace*{-.2\baselineskip}

\begin{figure}[t]
\centering
\includegraphics[width=0.8\linewidth]{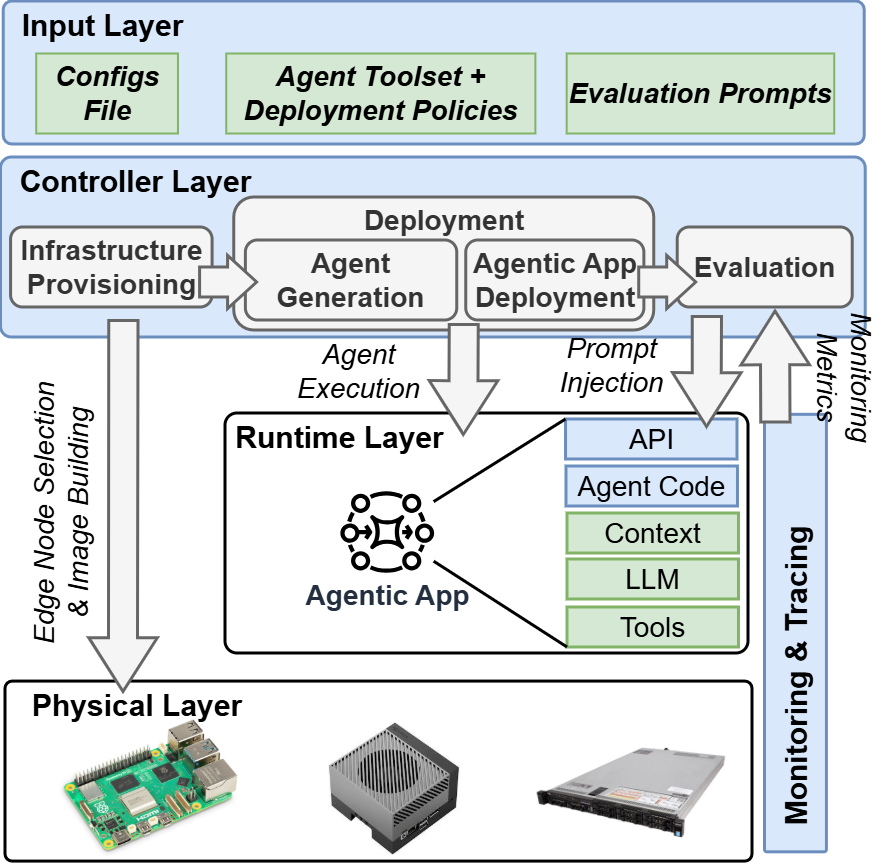}
    \vspace*{-.5\baselineskip}
\caption{AgentWare Architecture} \label{fig:architecture}
    \vspace*{-2\baselineskip}
\end{figure}

\noindent To materialize the AWOps pipeline, AgentWare adopts a multi-layer architecture comprising four layers: \textit{Input}, \textit{Control}, \textit{Runtime}, and \textit{Physical}, as illustrated in Fig.~\ref{fig:architecture}.
The \textit{Input Layer} serves as the user-facing interface of the framework. It receives the agent implementation, deployment configuration, and optional evaluation artifacts, validates and transforms them into a unified representation that drives the subsequent stages of the AWOps pipeline.
The \textit{Control Layer} orchestrates the execution of the entire lifecycle, coordinating infrastructure provisioning, runtime preparation, application deployment, observability configuration, and evaluation. Acting as the central coordination component, it translates high-level user specifications into actions executed across the underlying infrastructure.
In addition, the \textit{Physical Layer} represents the heterogeneous Edge-to-Cloud infrastructure on which applications are deployed. It encompasses edge devices, fog nodes, and cloud resources, each providing different computational capabilities, hardware architectures, and resource characteristics. This layer forms the execution substrate upon which the framework operates.
Lastly, the \textit{Runtime Layer} hosts the execution environment of the deployed agentic application. It contains the distributed agent components, supporting execution services, and observability infrastructure required for application operation and monitoring. The Runtime Layer abstracts infrastructure differences to provide uniform execution across heterogeneous resources while exposing telemetry and traces for evaluation.





\noindent \textbf{Input Specification: }
AgentWare is driven by three user-provided artifacts that collectively define the agentic application, its deployment configuration, and its evaluation process, enabling the framework to automate provisioning, deployment, observability, and evaluation while minimizing manual intervention.
The first artifact is the \textit{Agentic Toolset}, implemented as a Python module containing the agent's tools decorated with a description for each tool, the required arguments, the return value and the preferred physical-node placement
as illustrated in Fig. \ref{lst:user_input}. These tools constitute the functional capabilities exposed to the agent and may be accompanied by supporting artifacts such as datasets, model files, vector databases, or other runtime resources required by the tool implementations.

\noindent The second artifact is a declarative \textit{configuration file} in YAML format that specifies the deployment environment and runtime settings of the AWOps pipeline. As illustrated in Fig.~\ref{lst:config}, the configuration includes information about the available physical nodes, including their access details and a logical resource type used by the framework for the agent's tools placement. It also defines the LLM configuration, including the selected model and the physical node on which the model service will be deployed. At this stage, we envision integrating LLMFit~\footnote{\url{llmfit.org}} into the deployment stage so that AgentWare can automatically assess whether a user-selected LLM fits the computational and memory constraints of the target device before deployment. Furthermore, the configuration specifies the software dependencies required by the user-provided code, enabling the framework to automatically prepare a compatible execution environment during deployment. 
The third artifact is an optional \textit{prompts file} containing a collection of evaluation prompts. During the evaluation stage, the framework automatically submits these prompts to the deployed agent one by one, collecting the generated responses, execution traces, and infrastructure telemetry associated with each execution. When semantic evaluation is enabled through the configuration file, the collected execution artifacts are subsequently processed by the automated \textit{LLM-as-a-Judge} workflow, enabling reproducible assessment of agent behavior across various configurations without manual inspection of individual executions.

\lstset{
  language=Python,
  basicstyle=\ttfamily\scriptsize,
  keywordstyle=\color{blue}\bfseries,
  stringstyle=\color{blue},
  commentstyle=\color{gray}\itshape,
  showstringspaces=false,
  frame=none,
  breaklines=true,
  aboveskip=2pt,
  belowskip=2pt
}
\begin{figure}[!t]
\begin{minipage}[t]{0.48\columnwidth}
\begin{lstlisting}[language=Python]
nodes:
- name: "edge_server_1"
  ip: "10.16.27.29"
  username: "username"
  password: "password"
  resource_type: "edge_server"
- name: "edge_device_1"
  ip: "10.16.24.186"
  username: "username"
  password: "password"
  resource_type: "edge_node"
llm:
  host: "10.16.27.29"
  port: 11434
  model: "qwen3:8b"
\end{lstlisting}
\end{minipage}
\hfill
\begin{minipage}[t]{0.48\columnwidth}
\begin{lstlisting}[language=Python]
ssh_credentials: 
  username: "username"
  password: "password"
packages:
- easyocr
- python-Levenshtein
registry:
  url: "URL"
  username: "username"
  password: "password"
evaluation:
  provider: "openai"
  model: "gpt-4o"
  api_key: "OpenAI API key"
\end{lstlisting}
\end{minipage}
\vspace*{-.5\baselineskip}
\caption{Configuration Example.}
\label{lst:config}
\vspace*{-1.3\baselineskip}
\end{figure}

\noindent \textbf{Infrastructure Provisioning}
The Infrastructure Provisioning stage prepares the heterogeneous Edge-to-Cloud infrastructure for executing the agentic application (See Fig. \ref{fig:architecture}). Using the deployment configuration provided by the user in the input config file, AgentWare discovers the available nodes and their underlying hardware architectures. Based on this information, the framework automatically generates architecture-compatible execution artifacts (Docker Images) containing the software dependencies required by the agentic application and distributes them across the deployment. As a result, each node receives a runtime environment tailored to its hardware platform, abstracting architectural differences from the developer and enabling consistent execution across heterogeneous infrastructures. This automation creates an execution foundation for the subsequent deployment and evaluation stages.

\begin{figure}[t]
\centering
\includegraphics[width=0.8\linewidth]{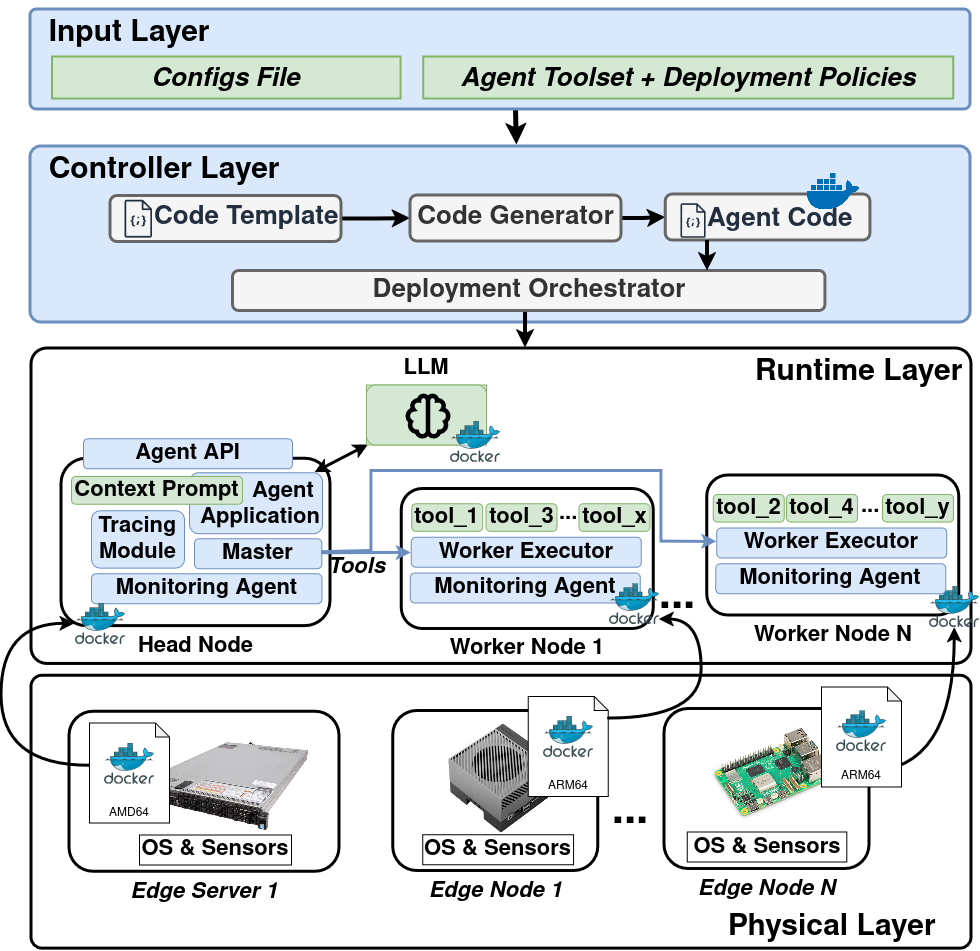}
\caption{Runtime \& Agentic Application Deployment} \label{fig:deployment}
\vspace*{-1.7\baselineskip}
\end{figure}

\noindent \textbf{Agent Generation \& Deployment}
This stage is responsible for generating the Runtime Layer (See Fig. \ref{fig:architecture}). At the beginning, AgentWare parses the user-provided configuration and the tools implementation files, as shown in Fig. \ref{fig:deployment}. 
During the \textit{Agent Generation} phase, the parsed artifacts are forwarded to the \textit{Code Generator}, which transforms the user-defined implementation into a distributed agentic application. Since existing agentic frameworks do not natively support deployment across heterogeneous Edge-to-Cloud infrastructures, AgentWare combines Smolagents~\footnote{\url{https://huggingface.co/docs/smolagents}} and Ray~\footnote{\url{http://www.ray.io}} to achieve this. We selected Smolagents because it provides modular, native-Python abstractions for agent components,
enabling AgentWare to automatically compose agents while preserving the underlying reasoning--action loop. Ray was selected because it provides task- and actor-based distributed execution primitives, enabling tool-level scheduling across heterogeneous resources. Unlike container orchestrators that primarily manage service placement, Ray allows AgentWare to dynamically execute individual tool invocations on specific Edge or Cloud nodes according to user-defined placement policies.
For the translation process, the \textit{Code Generator} uses Python's Abstract Syntax Tree (AST) representation to parse the user-defined tool functions and automatically transform them into framework-compatible implementations. For each user-defined tool (Fig.~\ref{lst:user_input}), the generator creates two complementary functions (Fig.~\ref{lst:tool_code_transformation}): a Ray-compatible remote function that encapsulates the original implementation and placement constraints, with the value 1 representing one available unit (physical device) of the defined resource type for the tool to be executed, and a Smolagents-compatible wrapper that exposes the tool metadata to the agent while delegating execution to the corresponding Ray function. During this transformation, AgentWare automatically injects tracing instrumentation and incorporates the selected LLM backend and optional context prompt template into the generated application code.
\lstdefinestyle{codeblock}{
language=Python,
numbers=left,
numberstyle=\tiny\color{black},
stepnumber=1,
firstnumber=1,
xleftmargin=2em,
breaklines=true,
columns=fullflexible
}
\begin{figure}[t!]
\centering
\begin{lstlisting}[language=Python, style=codeblock]
@agent_ware.tool(
description="Search for book information using web search",
args={
    "query": "Search query string about books or authors"
},
returns="Dictionary containing search query and top 5 search results",
resource="edge_device"
)
def search_book_info(query: str):
...  # User's implementation code
\end{lstlisting}
\vspace*{-.5\baselineskip}
\caption{User Input Code}
\label{lst:user_input}
\vspace*{-1.\baselineskip}
\end{figure}

\begin{figure}[t!]
\centering
\begin{lstlisting}[language=Python, style=codeblock]
@ray.remote(resources={'edge_device':1})
def search_book_info_r(query: str):
... # User's implementation code

@smolagents.tool
@tracer.tool(name='search_book_info')
def search_book_info(query: str):
"""
Search for book information using web search
Args:
   query: Search query string about books or authors
Returns:
   Dictionary of search query and top 5 search results"""
return ray.get(search_book_info_r.remote(query))
\end{lstlisting}
\vspace*{-.5\baselineskip}
\caption{System Output Code}
\label{lst:tool_code_transformation}
\vspace*{-1.5\baselineskip}
\end{figure}

Following code generation, the \textit{Deployment Orchestrator} materializes the runtime environment using the Docker images produced during provisioning (See Fig. \ref{fig:deployment}). It deploys the generated application to a Ray cluster consisting of a head node responsible for coordination and scheduling, and a set of worker nodes that execute tools according to the placement policies specified by the user. Depending on the deployment configuration, the selected LLM can be deployed on a designated infrastructure node or accessed through an external cloud endpoint.
To establish observability, AgentWare automatically deploys a monitoring stack that collects infrastructure-level telemetry, including CPU, memory, network, and energy metrics, together with agent-level traces capturing prompts, LLM calls, tool invocations, and execution timings. By combining these two perspectives, the framework provides the cross-layer observability required for evaluation and optimization.
Once deployed, the generated application exposes a unified API through which requests are submitted. The agent reasons over the request using the configured LLM, invokes the required tools through Ray, and returns the final response to the user.

\noindent \textbf{Monitoring Stack}
\label{subsec:monitoring_stack} 
Throughout execution, monitoring and tracing services continuously collect the telemetry and execution data that will later be consumed during the evaluation stage.
The current implementation uses containerized Netdata, a lightweight service that continuously gathers system metrics via host OS reporting mechanisms (e.g., cgroups, pseudofiles). The monitoring module also includes a custom probe for per-node power consumption, polling smart-plug APIs via HTTP and server PDUs via SNMP, or using software estimators such as RAPL for newer Intel CPU architectures. A centralized \textit{Monitoring Storage} component, implemented with Prometheus, periodically retrieves these metrics through the Netdata API, with configurations automatically generated by the orchestrator.
Beyond resource monitoring, agentic systems require agent-specific performance indicators such as tool-execution duration, end-to-end latency, and LLM tool-call accuracy. These are captured through a \textit{Tracing Module} based on Arize Phoenix, an open-source observability platform. Running on the head node and leveraging OpenTelemetry instrumentation, it records detailed agent events in the form of traces and spans, including tool invocations and LLM request timings to provide end-to-end visibility into agent behavior.

\begin{figure}[t]
\centering
\includegraphics[width=0.55\linewidth]{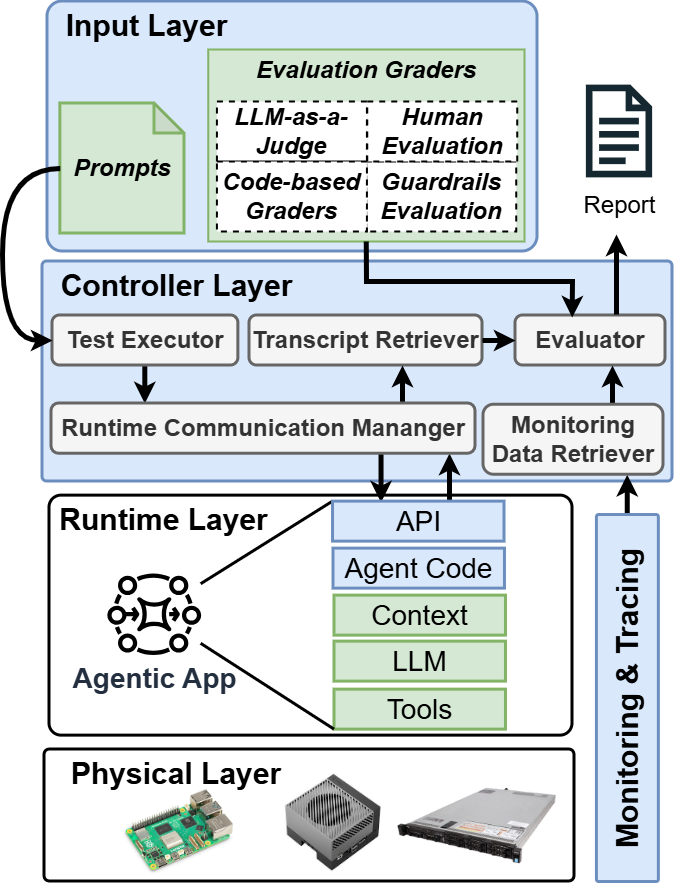}
\caption{Evaluation Pipeline} \label{fig:evaluation}
\vspace*{-1.8\baselineskip}
\end{figure}

\noindent \textbf{Evaluation}
The evaluation process (Fig.~\ref{fig:evaluation}) is initiated by the \textit{Test Executor}, which reads the prompts file and submits each prompt to the deployed agent through its API. For every execution, the framework records the submitted prompt, generated response, and execution timestamps, 
while concurrently collecting the corresponding execution traces and monitoring data.
After workload execution completes, the \textit{Evaluator} aggregates the collected telemetry and computes deployment-level metrics, including latency, resource utilization, and energy consumption. Infrastructure measurements are correlated with the execution window of each prompt, enabling attribution of resource usage and energy consumption to individual agent executions.
To support semantic evaluation, the Evaluator groups all spans belonging to the same trace identifier and reconstructs the corresponding execution path. From each trace, the framework extracts three artifacts: (i) the original user prompt, (ii) the tool definitions available to the agent, and (iii) the sequence of tool calls performed during execution of each prompt. All three artifacts are subsequently injected into a predefined \textit{LLM-as-a-Judge} template provided by Phoenix.
Using the evaluation configuration provided by the user, AgentWare automatically invokes the selected judge model, either locally or through a cloud service, and obtains a semantic correctness score together with a textual justification. 
Human-based assessment remains possible through the generated reports, which expose the prompt, response, execution transcript, telemetry metrics, and evaluation results for each execution. Finally, the Evaluator compiles all metrics into a unified report covering correctness, latency, resource use, and energy consumption, enabling comparison across deployment configurations, placement strategies, and LLMs.
As future work, we envision extending the evaluation stage with an action-level guardrail engine and code-based graders that consume execution traces to enforce configurable safety policies before tool invocation and verify user expected tool usage afterwards, enabling reproducible real-time evaluation.


\vspace*{-.2\baselineskip}
\section{Evaluation}
\vspace*{-.2\baselineskip}

AI coding assistants such as Codex or Claude can substantially accelerate the implementation of agent tools, and we used them to develop the tools of our book-assistant application. Then we followed AgentWare's programming model, using the deterministic configuration-driven pipeline to automate the lifecycle of the developed agentic application. We first validated AgentWare on a simple single-node "Hello World" agent before applying it to the representative distributed multi-tool application that combines image processing, retrieval, and external service interaction.
To quantify the automation benefits provided by AgentWare, we analyze the development effort required for the evaluated application. 
The book-assistant agent required only 177 lines of user-defined tool code and approximately 60 configuration lines, with the majority corresponding to framework defaults that can remain unchanged. From these inputs, AgentWare generated a 445-line deployment-ready application integrating distributed execution, tracing instrumentation, and runtime logic, while automatically building executable artifacts, deploying services, collecting traces, and executing the evaluation workflow. Without AgentWare, developers would manually implement distributed execution, deployment, monitoring, trace processing, and evaluation. AgentWare therefore reduces development effort and integration errors compared with manual lifecycle orchestration or relying solely on coding assistants.

\begin{figure}[t]
\centering
\includegraphics[width=0.98\linewidth]{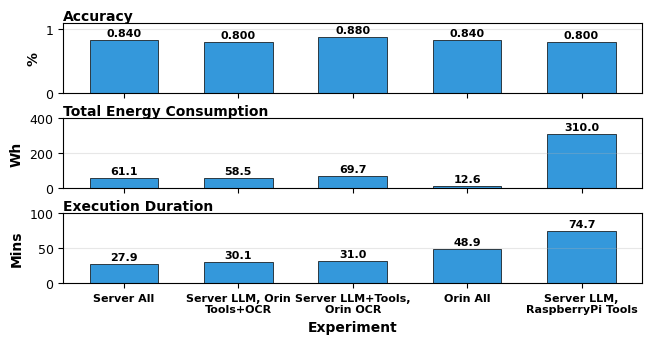}
\vspace*{-0.8\baselineskip}
\caption{Deployment Evaluation Metrics}
\label{fig:deployment_results}
\vspace*{-1.7\baselineskip}
\end{figure}

The implemented agent comprises three tools: (i) an OCR service (\texttt{process\_book\_cover}) based on EasyOCR 
that extracts textual information from book covers to support image-based identification; (ii) a retrieval service (\texttt{suggest\_similar\_books}) that uses fuzzy string matching
to recommend similar books based on the authors' similarity, and (iii) a web search service (\texttt{search\_book\_info}) that retrieves book information from external online sources. 


\noindent \textbf{Testbed Description:} 
The evaluation is conducted on a heterogeneous Edge-to-Cloud testbed comprising an Edge Server powered by a PDU, an NVIDIA Jetson AGX Orin, and a Raspberry Pi 4 powered by smart plugs. The Edge Server, representing the Near-Edge layer, is equipped with an Intel Xeon Gold 6230 CPU, an NVIDIA Tesla T4 GPU, and 96\,GB RAM. The Jetson AGX Orin represents a high-performance edge device featuring a 12-core Arm Cortex CPU, an NVIDIA Ampere GPU with 64 Tensor Cores, and 64\,GB memory. To further capture edge heterogeneity, we include a Raspberry Pi 4 equipped with a quad-core Arm Cortex-A72 CPU and 8\,GB RAM. These devices were intentionally selected to cover cloud-class GPU acceleration, high-performance edge acceleration, and resource-constrained edge execution.
For the evaluation, we designed a workload of 10 representative prompts with increasing reasoning complexity, and executed each prompt 5 times under every deployment configuration, yielding 50 executions per experiment. This repetition accounts for the stochastic nature of agentic systems, where LLM reasoning and tool-selection decisions may vary across executions even under fixed settings. The reported accuracy corresponds to the aggregate correctness across all repeated executions, capturing variations in agent behavior and providing a more reliable estimate than single-run evaluation.
GPU acceleration is enabled on both the Edge Server and Jetson AGX Orin whenever supported by the deployed services. Unless otherwise stated, we use Gemma 2 (9B) as the primary LLM with temperature set to 0, while semantic correctness is assessed through AgentWare's automated \textit{LLM-as-a-Judge} workflow using the GPT-4o model.

\begin{figure}[t]
\scriptsize

\textbf{Q1}: I have this book cover image at \texttt{bookcover1.jpg}. Can you identify the book and then suggest 3 other books by the same author?

\textbf{Q2}:  Identify the book in this image \texttt{bookcover1.jpg} and then search for information about the author's other works and their publication timeline.

\textbf{Q3}: Search for info about \textit{Dune} by Frank Herbert including its rating, then suggest me 4 similar books by the same author.

\textbf{Q4}: I found this book cover image \texttt{bookcover2.jpg}. Please identify it, search for critical reviews and awards it received, then suggest similar books by the same author.

\textbf{Q5}: Compare the publication years and critical reception of \textit{The Shining} by Stephen King and \textit{Rebecca} by Daphne du Maurier. Which one was published first and received better reviews?

\caption{Example Evaluation Prompts}
\vspace*{-0.5\baselineskip}
\label{fig:prompts}
\vspace*{-2\baselineskip}
\end{figure}




\vspace{-.5\baselineskip}
\subsection{Results}

A set of different scenarios are evaluated that a developer is likely to test in an Edge-enabled agentic application, namely, (i) different placement policies for agent services; (ii) the effects on accuracy and performance of different LLMs; (iii) the impact of changing the parameters of the application; and (iv) the observability in terms of per-query evaluation. 

\subsubsection{Evaluation of Service Placement}

This experiment evaluates the impact of service placement on accuracy, execution time, and energy consumption under five deployment configurations: (i) all services on the server, (ii) all services on the Orin, (iii) OCR on the Edge device with the remaining services on the server, (iv) the LLM on the server with all tools on the Orin, (v) the LLM on the server and all the tools on the Raspberry Pi 4. The application logic and LLM remain unchanged across deployments.
Fig.~\ref{fig:deployment_results} shows that accuracy remains relatively stable across deployments (0.80--0.88), with the small variations attributable to the inherent non-deterministic reasoning behavior of the same LLM used throughout the experiment, despite using deterministic decoding settings (temperature=0).
In contrast, deployment placement significantly affects the execution time and energy consumption. Hosting all tools on the Raspberry Pi requires the longest completion time ($\sim$75 minutes), followed by the 
Orin-only deployment ($\sim$49 minutes), whereas the rest of the deployments complete in 28--30 minutes. However, the Orin-only deployment consumes substantially less energy (12.6 Wh) compared to server-based deployments (58.5--310 Wh), reflecting the lower power profile of the edge device. On the other hand, the deployment involving the Raspberry Pi consumes the highest amount of energy due to the substantial amount of time required by the edge device to execute the OCR algorithm. At the same time, the server runs idle until further LLM requests are received.
These results highlight the trade-off between performance and sustainability. While server-based deployments reduce execution time, edge-only execution achieves up to 5$\times$ lower energy consumption, while accuracy is strongly dependent on the LLM behavior.


\noindent\textit{\textbf{Key Takeaway:} Service placement affects performance and energy consumption. Executing the whole application on the Orin provides substantial energy savings while preserving similar accuracy, at the cost of increased execution time.}

\subsubsection{Effect of LLM Choice}

\begin{figure}[t]
\centering
\includegraphics[width=0.98\linewidth]{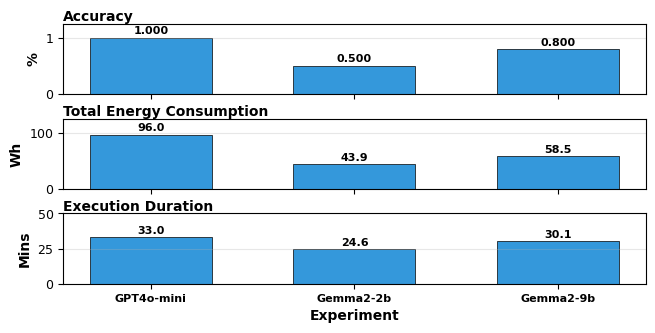}
\vspace*{-0.8\baselineskip}
\caption{Evaluation Metrics for different LLMs}
\label{fig:llm_results}
\vspace*{-1.5\baselineskip}
\end{figure}

This experiment evaluates the impact of LLM selection on agent accuracy, execution time, and energy consumption. The LLM is deployed on the server while the Orin executes the tools. Using the same set of 10 prompts, we compare: (i) Gemma 2 (9B) deployed locally, (ii) Gemma 2 (2B) deployed locally, and (iii) GPT-4o mini accessed through the OpenAI API. 
Fig.~\ref{fig:llm_results} shows a clear trade-off between accuracy, execution time, and energy consumption. Accuracy increases from 0.5 for Gemma 2 (2B) to 0.8 for Gemma 2 (9B), reaching 1.0 for GPT-4o mini. In terms of execution time, the smallest model is the fastest (24.6 min), followed by Gemma 2 (9B) (30 min) and GPT-4o mini (33 min).
For GPT-4o mini, only the Orin-side energy consumption can be directly measured (approximately 5 Wh), while the LLM executes remotely through the OpenAI API. The corresponding API cost for the evaluation workload was \$0.20. To estimate the energy consumed by the remote model, we apply the methodology of~\cite{jegham-2025}, obtaining an estimated total of 96 Wh, including both local and remote execution. For the local models, total energy consumption was 43.9 Wh for Gemma 2 (2B) and 58.5 Wh for Gemma 2 (9B). The higher consumption of the larger model is primarily attributed to its longer inference time, which increases both Edge device utilization and the server's idle power consumption.

\noindent\textit{\textbf{Key Takeaway:} LLM selection is a major determinant of agent quality and cost. GPT-4o mini achieves the highest accuracy, whereas smaller models offer lower execution time and energy consumption at the expense of reduced reasoning accuracy.}

\begin{figure}[t]
\centering
\includegraphics[width=0.98\linewidth]{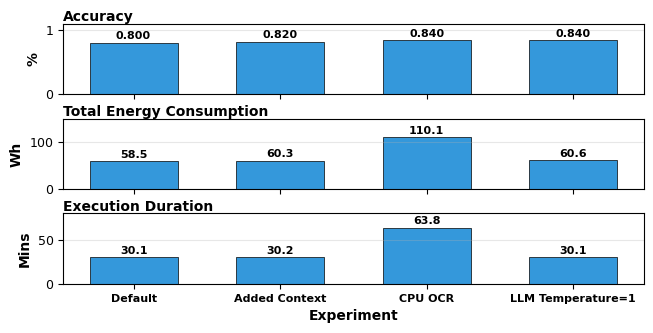}
\vspace*{-0.8\baselineskip}
\caption{Application Parameters Impact on Evaluation Metrics}
\label{fig:parameter_results}
\vspace*{-1.7\baselineskip}
\end{figure}

\begin{figure*}[!t]
\centering
\includegraphics[width=.9\linewidth, trim={0 .3cm 0 0.2cm}, clip]{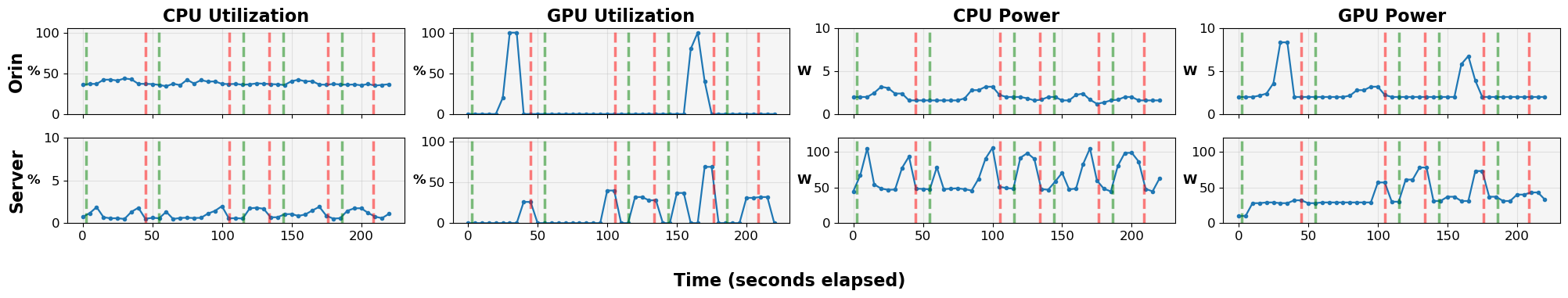}
  \vspace{-.3\baselineskip}
\caption{Utilization Metrics} 
\label{fig:utilization_metrics}
 \vspace{-1.7\baselineskip}
\end{figure*}

\subsubsection{Impact of Parameter Tuning}
This experiment evaluates the impact of application-level parameterization on agent behavior. Specifically, we vary: (i) the inclusion of additional application instructions template during reasoning, (ii) the execution of the OCR module on the Orin CPU instead of using GPU acceleration, and (iii) the LLM decoding temperature by setting it to 1.
\noindent The results are summarized in Fig.~\ref{fig:parameter_results}. Accuracy remains relatively stable across all configurations (0.80--0.84), suggesting that the agent is robust to these parameter changes. Switching the LLM temperature to 1 resulted in only minor accuracy variations, likely because Smolagents constrains tool selection through structured outputs, limiting LLM stochastic token generation on the execution workflow. Similarly, providing additional application context yields negligible improvement, indicating that the default agent instructions and tool descriptions already provide sufficient information for decision making.
\noindent In terms of performance and energy consumption, most parameter changes introduce little overhead. The main exception is executing the OCR module on the CPU, increasing the total execution time to 63 minutes and energy consumption to 110 Wh, nearly doubling both metrics compared to the GPU-accelerated execution. 

\noindent\textit{\textbf{Key Takeaway:} App-level parameterization has limited impact on accuracy in this case. However, implementation choices for compute-intensive tools, such as CPU vs GPU execution, can significantly affect both performance and energy consumption.}

\subsubsection{Per-prompt Utilization Metrics Analysis}


Finally, we evaluate the observability capabilities of AgentWare per prompt. Fig.~\ref{fig:utilization_metrics} presents resource-utilization metrics for the Orin and Server during the single execution of the 5 prompts presented in Fig. \ref{fig:prompts}. The deployment places the LLM on the server and all tools, including OCR, on the Orin. The green and red vertical lines indicate the start and end timestamps of each prompt.
Fig.~\ref{fig:utilization_metrics} shows a clear correlation between tool execution and resource utilization. On the Orin, GPU utilization and power consumption increase whenever OCR is invoked (Q1, Q4), while lightweight tool invocations produce only minor CPU activity. One repeated OCR request, as in Q2, exhibits lower-than-expected utilization, suggesting the presence of a possible caching effect when processing previously analyzed book covers.
On the server, GPU utilization primarily occurs during the final response generation stage, where the LLM produces the answer returned to the user. Although server CPU utilization remains low, even modest activity is sufficient to transition the processor from an idle to an active power state, resulting in a noticeable increase in CPU power consumption.
These observations show how AgentWare correlates agent-level execution traces with infrastructure-level telemetry, enabling attribution of resource utilization and energy consumption to individual prompts and tool invocations.

\noindent\textit{\textbf{Key Takeaway:} AgentWare enables fine-grained resource attribution, revealing patterns such as LLM-driven server GPU activity, caching effects, and CPU overheads. }
    \vspace{-0.5\baselineskip}
\section{Related Work}
         \vspace{-0.2\baselineskip}





Recent work has explored deploying AI and LLMs at the network edge to reduce latency, improve privacy, and support resource-constrained applications. Lin et al.~\cite{lin-2025} discuss the challenges of executing LLMs across heterogeneous edge infrastructures, while ~\cite{kasioulis-2024} investigates performance and energy trade-offs of AI inference on edge accelerators, highlighting the importance of sustainability-aware deployment decisions. Gupta et al.~\cite{gupta-2025} propose an LLM-based framework that automatically configures AWS EC2 instances, performs environment setup, and deploys web applications with minimal human intervention. Several frameworks have emerged to simplify agentic application development~\cite{wu-2024, fourney-2024}, providing abstractions for agent construction, multi-agent coordination, tool integration, and workflow orchestration.
While these frameworks facilitate the development of agentic systems, they largely assume a pre-existing execution environment and provide limited support for infrastructure-aware deployment across the Edge-to-Cloud. In contrast, AgentWare extends beyond agent development by automating the deployment, orchestration, monitoring, and evaluation of agentic applications across Edge-to-Cloud infrastructures.
Lastly, recent efforts have also focused on improving the observability and evaluation of agentic systems. Moshkovich et al.~\cite{moshkovich-2025} propose an observability-driven framework that captures detailed agent telemetry, while Dong et al.~\cite{dong-2024} define a taxonomy of observability requirements for LLM-enabled agents, including reasoning traces, tool usage, and performance monitoring. Building upon these ideas, AgentWare integrates observability with deployment automation, enabling unified collection of application traces and infrastructure metrics to support automated semantic evaluation and comparative analysis of deployments. 
Overall, existing research focuses on (i) agent development frameworks, (ii) EdgeAI and LLM deployment, or (iii) observability and evaluation of agentic systems in isolation. AgentWare bridges these domains by integrating lifecycle automation, observability, and evaluation into a unified AgenticOps workflow for Edge-to-Cloud environments.

\section{Conclusion}
In this paper, we presented AgentWare, an AgenticOps framework for automating the lifecycle of LLM-enabled agentic applications across the Edge-to-Cloud continuum. AgentWare integrates deployment automation, distributed execution, observability, and evaluation, enabling systematic comparison of correctness, performance, resource usage, and energy consumption. 
Our \textit{Future work} will extend the evaluation to multiple LLMs and real-world applications, while incorporating stronger baselines for effort reduction, statistical significance and variance analysis, and scalability experiments. We  also plan to improve correctness assessment through validated and deterministic evaluation methods, to complement LLM-as-a-Judge and provide more robust and reproducible results.



 
\vspace{.3\baselineskip}
\scriptsize{\noindent\textbf{Acknowledgments:} This work was 
supported by “Pharos-CY: Accelerating Trustworthy AI Innovation in Cyprus”, funded by the EuroHPC JU (GA: 101263007), with support from the Horizon Europe and the Government of the Republic of Cyprus. ChatGPT was used for language refinement; all ideas are the authors’ own.}

\vspace{-0.2\baselineskip}
\bibliographystyle{IEEEtran}
\bibliography{bibliography}

\end{document}